\documentclass[12pt]{iopart}
\usepackage{iopams}
\usepackage{graphicx}
 
\begin{document}

\title[Coupling between guided modes of two parallel nanofibers]{Coupling between guided modes of two parallel nanofibers}

\author{Fam Le Kien, Lewis Ruks, S\'{i}le Nic Chormaic and Thomas Busch}

\address{Okinawa Institute of Science and Technology Graduate University, Onna, Okinawa 904-0495, Japan}

\ead{kienle.pham@oist.jp}
\vspace{10pt}
\begin{indented}
\item[]\today
\end{indented}

\begin{abstract}
We study the coupling between the fundamental guided modes of two identical parallel nanofibers analytically and numerically. We calculate the coefficients of directional coupling, butt coupling, and mode energy changes as functions of the fiber radius, the light wavelength, and the fiber separation distance. We show that, due to the symmetry of the system, a mode of a nanofiber with the principal quasilinear polarization aligned along the radial or tangential axis is coupled only to the mode with the same corresponding principal polarization of the other nanofiber. We find that the effects of the butt coupling and the mode energy changes on the power transfer are significant when the fiber radius is small, the light wavelength is large, or the fiber separation distance is small. We show that the power transfer coefficient may achieve a local maximum or become zero when the fiber radius, the light wavelength, or the fiber separation distance varies indicating the system could be used in metrology for distance or wavelength measurements.
\end{abstract}

\vspace{2pc}
\noindent{\it Keywords}: guided modes, nanofibers, mode coupling
%\submitto{\NJP}
\maketitle

\section{Introduction}

Optical nanofibers are vacuum-clad, ultrathin optical fibers \cite{TongNat03} that allow tightly radially confined light to propagate over a long distance (the range of several millimeters is typical) and to interact efficiently with nearby atoms, molecules, quantum dots, or nanoparticles \cite{review2016,review2017,Nayak2018}. 
Optical nanofibers have been widely studied for various applications ranging from nonlinear optics to nanophotonics, quantum optics, and quantum photonics \cite{TongNat03,review2016,review2017,Nayak2018,Lukin2020}. 
Some uses of optical nanofibers include couplers for whispering gallery resonators \cite{Vahala1, Wu1}, miniature Mach-Zehnder interferometers \cite{Tong1}, sensors \cite{Tong2,Zhu1}, particle manipulation \cite{Tkachenko2}, generation of atom traps \cite{onecolor,twocolor,Vetsch2010,Goban2012}, channeling of light from quantum emitters into nanofiber-guided modes \cite{cesium decay,Nayak2007,Nayak2008,Kumar2015,Stourm2020}, scattering of 
nanofiber-guided light by atoms \cite{absorption,Sague2007}, 
use of nanofiber-guided light for Rydberg atom generation \cite{Rajasree2020}, 
and quadrupole excitations \cite{quadrupole,Ray2020}.
 
The mutual interaction between two copropagating light fields in the guided modes of two parallel waveguides is very important for the construction of practical optical devices \cite{Snyder1983,Marcus1989,Okamoto2006,Tamir2013}. 
Photonic structures composed of coupled micro- and nanofibers have been proposed, fabricated, and investigated \cite{L12,Jiang2006,Sumetsky2008,L16,L11,Glorieux2019}. Such structures have been used to build miniaturized  ring interferometers and  knot resonators in the visible light wavelength range in recent work by Ding \textit{et al.} \cite{Glorieux2019}. They used the linear coupling theory \cite{Snyder1983,Marcus1989,Okamoto2006,Tamir2013,L4,L5,L6,Hardy1985,L7,Huang1990,L8,L9,Chremmos2006} to calculate the directional coupling coefficient for different fiber radii and different polarization angles. However, as in almost all work related to coupling of optical fibers, the coefficients for butt coupling and mode energy changes \cite{Snyder1983,Marcus1989,Okamoto2006} have been omitted despite the fact that these effects could be quite significant and can lead to dramatically different results in the case of  nanofibers due to the significant spatial spread and overlap of the modes.

The purpose of this paper is to present a systematic and complete treatment for the coupling between the fundamental guided modes of two identical parallel nanofibers. Using coupled mode theory, we show that the effects of the butt coupling and the mode energy changes on the power transfer are significant when the fiber radius is small, the light wavelength is large, or the fiber separation distance is small. 

The paper is organized as follows. In section \ref{sec:model} we describe the model and present the basic coupled mode equations. Section \ref{sec:coefficients} is devoted to the calculations of the coupling coefficients for the modes with the principal quasilinear polarizations.
In section \ref{sec:transfer}, we present the results of calculations for the power transfer and phase shift coefficients. Our conclusions are given in section \ref{sec:summary}.

\section{Two coupled parallel nanofibers and basic mode coupling equations
}
\label{sec:model}

%%%%%%%%%%%%%%%%%%%%%%% Figure 1
\begin{figure}[tbh]
	\begin{center}
		\includegraphics{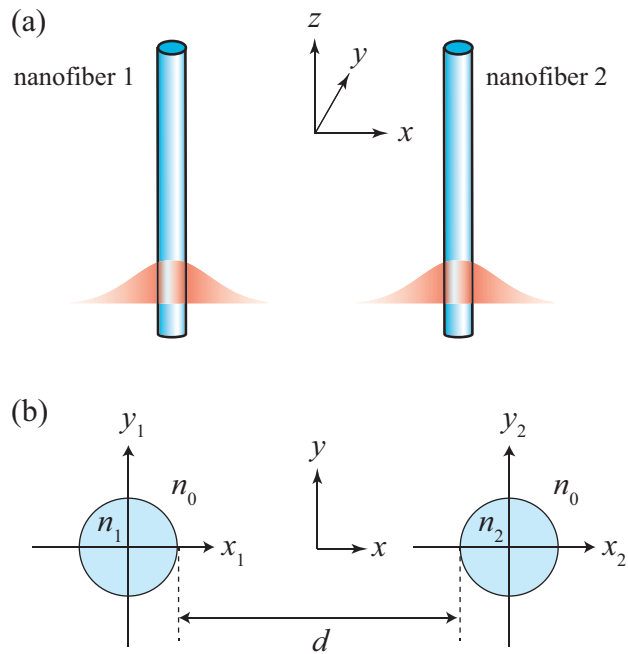}
	\end{center}
	\caption{(a) Directionally coupled nanofibers. (b) Spatial profile of the refractive index in the fiber transverse plane $(x,y)$. 
	}
	\label{fig1}
\end{figure}

We consider two identical vacuum-clad, ultrathin optical fibers aligned parallel to each other as shown in figure \ref{fig1}. We label the fibers by the indices $j=1,2$. 
Each fiber $j$ is a dielectric cylinder of radius $a$ and refractive index $n_1=n_2>1$ and is surrounded by an infinite background vacuum or air medium of refractive index $n_0=1$. We are interested in vacuum-clad, silica-core fibers with diameters in the range of hundreds of nanometers. Such ultrathin optical fibers are usually called nanofibers \cite{TongNat03,review2016,review2017,Nayak2018}. They can support one or a few modes depending on the fiber size parameter $V=ka\sqrt{n_j^2-n_0^2}$, where $k$ is the free-space wave number of light. 
For $V<V_{\mathrm{cutoff}}$ where $V_{\mathrm{cutoff}}\simeq2.405$ is the cutoff parameter, the fiber can support only the fundamental mode  HE$_{11}$, while for $V> V_{\mathrm{cutoff}}$, the fiber can support not only the fundamental mode but also higher-order modes \cite{TongNat03,Snyder1983,Marcus1989,Okamoto2006,higherorder1,higherorder2}.

We use the global Cartesian coordinate system $\{x,y,z\}$, where the axis $z$ is parallel to the fiber axes $z_1$ and $z_2$, the axis $x$ is perpendicular to the axis $z$ and connects the fiber centers, and the axis $y$ is perpendicular to the axes $x$ and $z$ (see figure \ref{fig1}). The cross-sectional plane of the fibers is the plane $xy$. 
We call the axes $x$ and $y$ the radial (horizontal) and tangential (vertical) axes, respectively, of the coupled fiber system [see figure \ref{fig1}(b)]. The corresponding polar coordinate system is denoted as $\{r,\varphi\}$. Without loss of generality, we assume that the fiber axes $z_1$ and $z_2$ are positioned symmetrically at the points $x=\pm(a+d/2)$ on the axis $x$, where $d$ is the separation distance between the fibers. For each individual fiber $j$, we use the local fiber-based Cartesian coordinate system $\{x_j,y_j,z_j\}$, where the origin is at the center of the respective fiber, the axis $x_j$ is aligned along $x$, and the axis $y_j$ is parallel to $y$ [see figure \ref{fig1}(b)]. We also use the corresponding local polar coordinate systems $\{r_j,\varphi_j\}$.

We study the coupling between two copropagating guided light fields of the same optical frequency $\omega$ in the fundamental modes HE$_{11}$ of the nanofibers. Without loss of generality, we assume that these guided fields propagate in the forward direction $+z$. Since the phase mismatch between counterpropagating modes is large, the coupling between them is weak and hence is neglected \cite{Snyder1983,Marcus1989,Okamoto2006}. For the same reason, we neglect the coupling between guided modes and radiation modes \cite{Snyder1983,Marcus1989,Okamoto2006}.  

The guided field in each nanofiber before mode coupling can be decomposed into a superposition of the orthogonal fundamental modes with the orthogonal polarizations $p_1$ and $p_2$.
Let $\mathbf{E}_{j}^{(p)}$ and $\mathbf{H}_{j}^{(p)}$ be the positive frequency parts of the electric and magnetic components of the fundamental mode HE$_{11}$ with the polarization $p=p_1,p_2$ of nanofiber $j=1,2$ before mode coupling \cite{Snyder1983,Marcus1989,Okamoto2006}. As the main approximation in the coupled mode theory, we assume that the field vector $(\mathbf{E},\mathbf{H})$ of the coupled nanofibers can be expressed as a superposition of the field vectors $(\mathbf{E}_j^{(p)},\mathbf{H}_j^{(p)})$ of the fundamental modes of the individual fibers, that is, 
\begin{equation}\label{a1}
\mathbf{E}=\sum_{jp}A_j^{(p)}\mathbf{E}_j^{(p)},\qquad
\mathbf{H}=\sum_{jp}A_j^{(p)}\mathbf{H}_j^{(p)}.
\end{equation}
Here, $A_j^{(p)}$ are $z$-dependent coefficients.
We separate the transverse and axial dependencies of $\mathbf{E}_j^{(p)}$ and $\mathbf{H}_j^{(p)}$ as 
\begin{equation}\label{a2}
\mathbf{E}_j^{(p)}=\boldsymbol{\mathcal{E}}_j^{(p)}(x,y)\rme^{\rmi\beta z},\qquad
\mathbf{H}_j^{(p)}=\boldsymbol{\mathcal{H}}_j^{(p)}(x,y)\rme^{\rmi\beta z},
\end{equation}
where $\beta$ is the longitudinal propagation constant of the fundamental guided modes of the individual fibers \cite{Snyder1983,Marcus1989,Okamoto2006}. The amplitudes $A_j^{(p)}$ are governed by the equations \cite{Snyder1983,Marcus1989,Okamoto2006}
\begin{equation}\label{a3}
\frac{\rmd A_j^{(p)}}{\rmd z}+\sum_{p'}c_{jj'}^{(pp')}\frac{\rmd A_{j'}^{(p')}}{\rmd z}
-\rmi\sum_{p'}\chi_j^{(pp')}A_j^{(p')}
-\rmi\sum_{p'}\kappa_{jj'}^{(pp')}A_{j'}^{(p')}=0,
\end{equation}
where $j'\not=j$, that is, $(j,j')=(1,2)$ or $(2,1)$, and
\begin{eqnarray}\label{a4}
\kappa_{jj'}^{(pp')}&=&\frac{\omega\epsilon_0\int \rmd\mathbf{r}\; (N^2-N_{j'}^2)
	(\boldsymbol{\mathcal{E}}_j^{(p)*}\cdot\boldsymbol{\mathcal{E}}_{j'}^{(p')})}
{2\int \rmd\mathbf{r}\; \mathrm{Re}[\boldsymbol{\mathcal{E}}_j^{(p)*}\times \boldsymbol{\mathcal{H}}_{j}^{(p)}]_z},
\nonumber\\
c_{jj'}^{(pp')}&=&\frac{\int \rmd\mathbf{r}\; 
	\{[\boldsymbol{\mathcal{E}}_j^{(p)*}\times\boldsymbol{\mathcal{H}}_{j'}^{(p')}]_z+
	[\boldsymbol{\mathcal{E}}_{j'}^{(p')}\times\boldsymbol{\mathcal{H}}_{j}^{(p)*}]_z\}
}
{2\int \rmd\mathbf{r}\; \mathrm{Re}[\boldsymbol{\mathcal{E}}_j^{(p)*}\times \boldsymbol{\mathcal{H}}_{j}^{(p)}]_z},
\nonumber\\
\chi_j^{(pp')}&=&\frac{\omega\epsilon_0\int \rmd\mathbf{r}\; (N^2-N_j^2)
	(\boldsymbol{\mathcal{E}}_j^{(p)*}\cdot\boldsymbol{\mathcal{E}}_{j}^{(p')})}
{2\int \rmd\mathbf{r}\; \mathrm{Re}[\boldsymbol{\mathcal{E}}_j^{(p)*}\times \boldsymbol{\mathcal{H}}_{j}^{(p)}]_z}
\end{eqnarray}
are the mode coupling coefficients. Here, $N_j$ is the refractive-index distribution of the space in the presence of fiber $j$ alone, $N$ is the refractive-index distribution of the space in the presence of the coupled fibers 1 and 2, and $\int \rmd\mathbf{r}=\int_{-\infty}^\infty\int_{-\infty}^\infty \rmd x \rmd y$ is the integral over the fiber transverse plane. Note that the integrals in the numerators of the expressions for $\kappa_{jj'}^{(pp')}$ and $\chi_j^{(pp')}$ are carried out in the cross-sectional areas of fibers $j$ and $j'$, respectively, where the factors $N^2-N_{j'}^2$ and $N^2-N_j^2$ are not zero, respectively. Meanwhile, the integral in the numerator of the expression for $c_{jj'}^{(pp')}$ is carried out in the full transverse plane $xy$. The integral that is present in all of the denominators is twice  the power and is also carried out in the full transverse plane $xy$.
The coefficients $\kappa_{jj'}^{(pp')}$ are the mode coupling coefficients of the directional coupler \cite{Snyder1983,Marcus1989,Okamoto2006}. The coefficients $c_{jj'}^{(pp')}$ represent the butt coupling (cross-power coupling) of the waveguides \cite{Okamoto2006}. The coefficients $\chi_j^{(pp')}$ characterize 
the self coupling resulting from the changes of the electromagnetic energies of the modes of fiber $j$ induced by the presence of the other perturbing fiber. We note that the directional coupling coefficients $\kappa_{jj'}^{(pp')}$ and the self-coupling (mode-energy-change) coefficients $\chi_j^{(pp')}$ have the dimension of inverse length, while the butt coupling coefficients $c_{jj'}^{(pp')}$ are dimensionless. The characteristic lengths $1/\kappa_{jj'}^{(pp')}$ and $1/\chi_j^{(pp')}$ determine the propagation distances required for the directional coupling and mode energy changes to be substantial. The dimensionless coefficients $c_{jj'}^{(pp')}$ characterize the efficiency of the butt coupling.

While butt coupling is usually interpreted as a direct light field transfer in a structure composed of two end-to-end joined fibers, where light exiting from a transmitting fiber is coupled to a receiving fiber, it can also occur in directional couplers \cite{Okamoto2006}. In the case of directional couplers, butt coupling means that a gradient of the envelope of the field of a fiber can excite the field of the other fiber and should not be ignored. 

The set of the coefficients $\chi_j^{(pp')}$ describes not only the shifts of the energies and propagation constants of the modes ($p=p'$) of fiber $j$ but also the coupling between the different modes ($p\not=p'$) of the fiber. This leads to splitting of the degenerate mode energies and propagation constants and also to mode mixing. Note that $\chi_j^{(pp')}=\kappa_{jj}^{(pp')}$.

When the local modes $p$ and $p'$ are normalized to have the same power, the
coefficients $c_{jj'}^{(pp')}$ and $\chi_j^{(pp')}$ satisfy the Hermitian conjugate
relationship, that is, $c_{jj'}^{(pp')}=c_{j'j}^{(p'p)*}$ and $\chi_j^{(pp')}=\chi_j^{(p'p)*}$. 
Meanwhile, the coefficients $\kappa_{jj'}^{(pp')}$ in general do not satisfy this relationship unless the waveguides are identical \cite{Snyder1983,Marcus1989,Okamoto2006}. 
We will show in the next section that, in the case of identical nanofibers as considered in this paper, the coefficients $\kappa_{jj'}^{(pp')}$ satisfy the Hermitian conjugate relationship, that is, $\kappa_{jj'}^{(pp')}=\kappa_{j'j}^{(p'p)*}$ \cite{Snyder1983,Marcus1989,Okamoto2006}. 

The eigenvectors of the coupled differential equations (\ref{a3}) represent the normal modes of the coupled fibers \cite{Snyder1983,Marcus1989,Okamoto2006}. The eigenvalues are the corrections to the propagation constants of the normal modes.
It is clear that the whole set of the coupling coefficients $\kappa_{jj'}^{(pp')}$, $c_{jj'}^{(pp')}$, and $\chi_{j}^{(pp')}$ determines the normal modes and their propagation constants.

The sets of the orthogonal modes with the polarization indices $p_1$ and $p_2$ of each nanofiber $j=1,2$ can, in principle, be arbitrary. Equations (\ref{a3}) and (\ref{a4}) indicate that each mode $p_1$ or $p_2$ of a nanofiber is, in general, coupled to both modes $p_1$ and $p_2$ of the other nanofiber. In addition, modes $p_1$ and $p_2$ of a nanofiber can also be coupled to each other due to the presence of the other nanofiber.

\section{Coupling coefficients for the modes with the principal quasilinear polarizations}
\label{sec:coefficients} 

Without loss of generality, we use the basis modes that are quasilinearly polarized \cite{Snyder1983,Marcus1989,Okamoto2006,Tong2004,fibermode,highorder} along the radial axis $x$ and the tangential axis $y$ and are labeled by the polarization index $p=\mathbf{x}$ and $\mathbf{y}$, respectively. 
We refer to the quasilinear polarizations along these main axes as the principal quasilinear polarizations.
We introduce the notations $\boldsymbol{\mathcal{E}}^{(p)}_j$ and $\boldsymbol{\mathcal{H}}^{(p)}_j$ for the envelopes of the electric and magnetic parts of the $p$-polarized forward fundamental mode of nanofiber $j$. 
For convenience and without loss of generality, we use the sets $(\boldsymbol{\mathcal{E}}^{(p)}_1,\boldsymbol{\mathcal{H}}^{(p)}_1)$
and $(\boldsymbol{\mathcal{E}}^{(p)}_2,\boldsymbol{\mathcal{H}}^{(p)}_2)$ that
are identical to each other in their corresponding local fiber-based coordinate systems,
that is, $\boldsymbol{\mathcal{E}}^{(p)}_1(r_1,\varphi_1)=\boldsymbol{\mathcal{E}}^{(p)}_2(r_2,\varphi_2)$ and
$\boldsymbol{\mathcal{H}}^{(p)}_1(r_1,\varphi_1)=\boldsymbol{\mathcal{H}}^{(p)}_2(r_2,\varphi_2)$ for $r_1=r_2$ and $\varphi_1=\varphi_2$. For convenience, we normalize the sets of the mode functions $(\boldsymbol{\mathcal{E}}^{(p)}_j,\boldsymbol{\mathcal{H}}^{(p)}_j)$ for different indices $j$ and $p$ to have the same power $P_0$. The explicit expressions for $\boldsymbol{\mathcal{E}}^{(p)}_j$ and $\boldsymbol{\mathcal{H}}^{(p)}_j$ are given in \ref{appendix}.

%%%%%%%%%%%%%%%%%%%%%%% Figure 2
\begin{figure}[tbh]
	\begin{center}
		\includegraphics{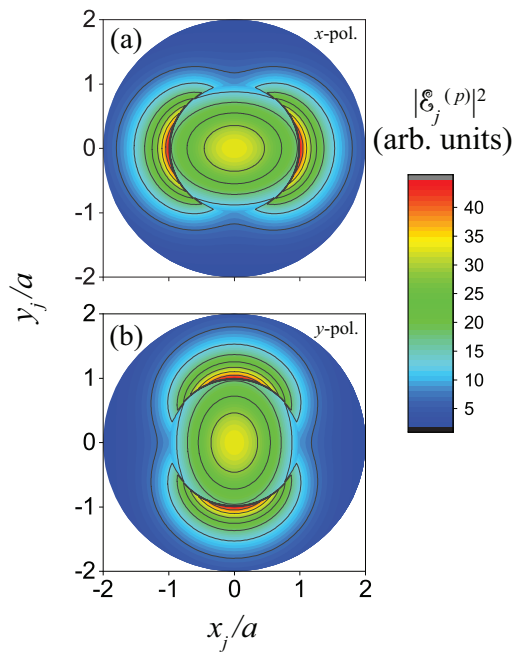}
	\end{center}
	\caption{Cross-sectional profiles of the electric intensity distributions 
		$|\boldsymbol{\mathcal{E}}_j^{(p)}|^2$ of the fields in the (a) $x$- and (b) $y$-polarized fundamental guided modes of nanofiber $j$. 
		The fiber radius is $a=200$ nm.
		The refractive indices of the silica core and the vacuum cladding are $n_j=1.45$ and $n_0=1$, respectively. The free-space wavelength of light is $\lambda=800$ nm. 
	}
	\label{fig2}
\end{figure}

We plot in figure \ref{fig2} the cross-sectional profiles of the electric intensity distributions $|\boldsymbol{\mathcal{E}}_j^{(p)}|^2$ of the fields in the $x$- and $y$-polarized fundamental guided modes of a single nanofiber $j$. 
The free-space wavelength of light is $\lambda=800$ nm, and the refractive indices of the fibers and the vacuum cladding are $n_1=n_2=1.45$ and $n_0=1$, respectively. 
The figure shows that the intensity distribution $|\boldsymbol{\mathcal{E}}_j^{(p)}|^2$ is symmetric with respect to the $x_j$ and $y_j$ axes. The symmetry of the intensity distribution is a consequence of the fact that the Cartesian components of the mode functions of the quasilinearly polarized modes are even or odd functions of the transverse coordinates $x_j$ and $y_j$ [see equations (\ref{b3}) and (\ref{b4})]. 

It follows from expressions (\ref{b3}) and (\ref{b4}) that,
when we perform the transformation $(x,y)\to(x,-y)$, that is, $(r_j,\varphi_j)\to(r_j,-\varphi_j)$, the field components are transformed as described by the formulas
\begin{eqnarray}\label{a5}
\mathcal{E}^{(\mathbf{x})}_{jx}&\to&\mathcal{E}^{(\mathbf{x})}_{jx},
\qquad
\mathcal{E}^{(\mathbf{x})}_{jy}\to-\mathcal{E}^{(\mathbf{x})}_{jy},
\qquad
\mathcal{E}^{(\mathbf{x})}_{jz}\to\mathcal{E}^{(\mathbf{x})}_{jz},
\nonumber\\
\mathcal{H}^{(\mathbf{x})}_{jx}&\to&-\mathcal{H}^{(\mathbf{x})}_{jx},
\qquad
\mathcal{H}^{(\mathbf{x})}_{jy}\to\mathcal{H}^{(\mathbf{x})}_{jy},
\qquad
\mathcal{H}^{(\mathbf{x})}_{jz}\to-\mathcal{H}^{(\mathbf{x})}_{jz},
\end{eqnarray}
and
\begin{eqnarray}\label{a6}
\mathcal{E}^{(\mathbf{y})}_{jx}&\to&-\mathcal{E}^{(\mathbf{y})}_{jx},
\qquad
\mathcal{E}^{(\mathbf{y})}_{jy}\to\mathcal{E}^{(\mathbf{y})}_{jy},
\qquad
\mathcal{E}^{(\mathbf{y})}_{jz}\to-\mathcal{E}^{(\mathbf{y})}_{jz},
\nonumber\\
\mathcal{H}^{(\mathbf{y})}_{jx}&\to&\mathcal{H}^{(\mathbf{y})}_{jx},
\qquad
\mathcal{H}^{(\mathbf{y})}_{jy}\to-\mathcal{H}^{(\mathbf{y})}_{jy},
\qquad
\mathcal{H}^{(\mathbf{y})}_{jz}\to\mathcal{H}^{(\mathbf{y})}_{jz},
\end{eqnarray}
which lead to
\begin{equation}\label{a7}
(\boldsymbol{\mathcal{E}}^{(\mathbf{x})*}_j\cdot\boldsymbol{\mathcal{E}}^{(\mathbf{y})}_{j'})\to -(\boldsymbol{\mathcal{E}}^{(\mathbf{x})*}_j\cdot\boldsymbol{\mathcal{E}}^{(\mathbf{y})}_{j'}),
\quad
[\boldsymbol{\mathcal{E}}_j^{(\mathbf{x})*}\times\boldsymbol{\mathcal{H}}^{(\mathbf{y})}_{j'}]_z\to
-[\boldsymbol{\mathcal{E}}_j^{(\mathbf{x})*}\times\boldsymbol{\mathcal{H}}^{(\mathbf{y})}_{j'}]_z.
\end{equation}
We use the above symmetry relations to calculate the integrals in equations (\ref{a4}). Then, we obtain
\begin{equation}\label{a8}
\kappa_{jj'}^{(pp')}=\kappa_{jj'}^{(pp)}\delta_{pp'},\qquad
c_{jj'}^{(pp')}=c_{jj'}^{(pp)}\delta_{pp'},\qquad
\chi_j^{(pp')}=\chi_j^{(pp)}\delta_{pp'}.
\end{equation} 
According to equations (\ref{a8}), the mode coupling coefficients $\kappa_{jj'}^{(pp')}$, $c_{jj'}^{(pp')}$, and $\chi_j^{(pp')}$ are zero for $(p,p')=(\mathbf{x},\mathbf{y})$ or $(\mathbf{y},\mathbf{x})$, that is, for the pair of orthogonal $x$- and $y$-polarized modes. 
Thus, the two polarizations are decoupled. This decoupling is a consequence of the symmetry of the two-fiber system about the principal $x$ and $y$ axes.
Note that the first relation in equations (\ref{a8}) is in agreement with \cite{Marcus1989}.

For an appropriate choice of the phase of the normalization constant for the mode functions, we have \cite{Snyder1983,Marcus1989,Okamoto2006,Tong2004,fibermode,highorder}
\begin{eqnarray}\label{a9}
&& e_r^*=-e_r,\qquad e_\varphi^*=e_\varphi,\qquad e_z^*=e_z,\nonumber\\ 
&& h_r^*=h_r,\qquad h_\varphi^*=-h_\varphi,\qquad h_z^*=-h_z.
\end{eqnarray}
When we use the above relations and expressions (\ref{b3}) and (\ref{b4}), 
we can show that 
\begin{eqnarray}\label{a10}
&&\mathcal{E}^{(p)*}_{jx}=-\mathcal{E}^{(p)}_{jx},\qquad 
\mathcal{E}^{(p)*}_{jy}=-\mathcal{E}^{(p)}_{jy},\qquad
\mathcal{E}^{(p)*}_{jz}=\mathcal{E}^{(p)}_{jz},\nonumber\\
&&\mathcal{H}^{(p)*}_{jx}=-\mathcal{H}^{(p)}_{jx},\qquad 
\mathcal{H}^{(p)*}_{jy}=-\mathcal{H}^{(p)}_{jy},\qquad
\mathcal{H}^{(p)*}_{jz}=\mathcal{H}^{(p)}_{jz}.\qquad
\end{eqnarray}
Then, equations (\ref{a4}) yield
\begin{equation}\label{a11}
\kappa_{jj'}^{(pp')}=\kappa_{jj'}^{(pp')*}, \qquad 
c_{jj'}^{(pp')}=c_{jj'}^{(pp')*}, \qquad 
\chi_j^{(pp')}=\chi_j^{(pp')*}.
\end{equation} 
Thus, for an appropriate choice of the phase of the normalization constant for the mode functions, the mode coupling coefficients 
$\kappa_{jj'}^{(pp')}$, $c_{jj'}^{(pp')}$, and $\chi_j^{(pp')}$ for the $x$- and $y$-polarized modes are real-valued coefficients.

It follows from expressions (\ref{b3}) and (\ref{b4}) that, in the central Cartesian coordinate system $\{x,y\}$ for the fiber transverse plane, we have the relations
\begin{eqnarray}\label{a12}
\mathcal{E}^{(p)}_{1x}(x,y)&=&\mathcal{E}^{(p)}_{2x}(-x,-y),
\nonumber\\
\mathcal{E}^{(p)}_{1y}(x,y)&=&\mathcal{E}^{(p)}_{2y}(-x,-y),
\nonumber\\
\mathcal{E}^{(p)}_{1z}(x,y)&=&-\mathcal{E}^{(p)}_{2z}(-x,-y),
\end{eqnarray}
and
\begin{eqnarray}\label{a13}
\mathcal{H}^{(p)}_{1x}(x,y)&=&\mathcal{H}^{(p)}_{2x}(-x,-y),
\nonumber\\
\mathcal{H}^{(p)}_{1y}(x,y)&=&\mathcal{H}^{(p)}_{2y}(-x,-y),
\nonumber\\
\mathcal{H}^{(p)}_{1z}(x,y)&=&-\mathcal{H}^{(p)}_{2z}(-x,-y).
\end{eqnarray}
Hence, we can show that
\begin{equation}\label{a14} 
\kappa_{12}^{(pp)}=\kappa_{21}^{(pp)}, \qquad 
c_{12}^{(pp)}=c_{21}^{(pp)}, \qquad 
\chi_1^{(pp)}=\chi_2^{(pp)}. 
\end{equation}
It follows from equations (\ref{a8}), (\ref{a11}), and (\ref{a14}) that all the mode coupling coefficients $\kappa_{jj'}^{(pp')}$, $c_{jj'}^{(pp')}$, and $\chi_{j}^{(pp')}$ satisfy the Hermitian conjugate relationship, that is, $\kappa_{jj'}^{(pp')}=\kappa_{j'j}^{(p'p)*}$, $c_{jj'}^{(pp')}=c_{j'j}^{(p'p)*}$, 
and $\chi_{j}^{(pp')}=\chi_{j}^{(p'p)*}$.

Due to the diagonal properties (\ref{a8}) and the symmetry properties (\ref{a14}), it is convenient to use the simplified notations $\kappa_p=\kappa_{12}^{(pp)}=\kappa_{21}^{(pp)}$, $c_p=c_{12}^{(pp)}=c_{21}^{(pp)}$, and $\chi_p=\chi_1^{(pp)}=\chi_2^{(pp)}$. In the following, we calculate the coupling coefficients $\kappa_p$, $c_p$, and $\chi_p$ for the pairs of quasilinearly polarized fundamental modes of the fibers with the same $x$ or $y$ principal polarization.
According to equations (\ref{a11}), these coefficients are real-valued.

%%%%%%%%%%%%%%%%%%%%%%% Figure 3
\begin{figure}[tbh]
	\begin{center}
		\includegraphics{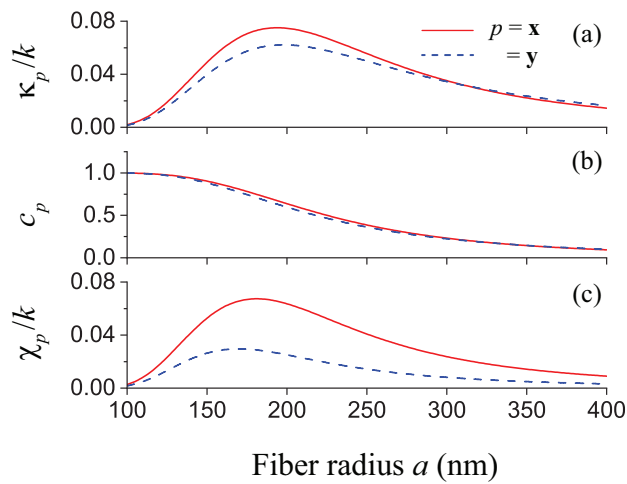}
	\end{center}
	\caption{Directional coupling coefficient $\kappa_p$ (a), butt coupling coefficient $c_p$ (b), and self-coupling coefficient $\chi_p$ as functions of the fiber radius $a$. 
		The coefficients $\kappa_p$ and $\chi_p$ are normalized to the wave number $k=2\pi/\lambda$ of light. The wavelength of light is $\lambda=800$ nm. 
		The separation distance between the fibers is $d=0$. The mode polarization is $p=\mathbf{x}$ (solid red lines) or $\mathbf{y}$ (dashed blue lines). Other parameters are as in figure \ref{fig2}.
	}
	\label{fig3}
\end{figure}

We plot in figure \ref{fig3} the directional coupling coefficient $\kappa_p$, the butt coupling coefficient $c_p$, and the self-coupling coefficient $\chi_p$ as functions of the fiber radius $a$. The wavelength of light is $\lambda=800$ nm and the separation distance between the fibers is $d=0$. The figure shows that the directional coupling coefficient $\kappa_p$ has a peak at almost the same point $a\simeq 195$ nm for both the $x$ and $y$ polarizations, and the self-coupling coefficient $\chi_p$ has a peak at $a\simeq 180$ and 170 nm for the $x$ and $y$ polarizations, respectively. Meanwhile, the butt coupling coefficient $c_p$ monotonically reduces with increasing fiber radius $a$. Note that the peak in the dependence of the coefficient $\kappa_p$ on $a$ was not observed in \cite{Glorieux2019} because calculation results were not presented for $a\leq 200$ nm. In the limit of large fiber radii $a\gg\lambda/2\pi$, the coefficients $\kappa_p$, $c_p$, and $\chi_p$ reduce to zero. This behavior is a consequence of the reduction of the overlap between the modes of different fibers in the limit of large $a$. In the limit of small fiber radii $a\ll\lambda/2\pi$, the coefficients $\kappa_p$ and $\chi_p$ reduce to zero and the coefficient $c_p$ increases to 1. The difference in the limiting value is due to the facts that the integrals in the numerators of the expressions for $\kappa_p$ and $\chi_p$ in equations (\ref{a4}) are carried out in the cross-sectional area of a fiber, while the integral in the numerator of the expression for $c_p$ is carried out in the full transverse plane $xy$ and, for a small fiber size parameter, the mode profiles extend far beyond the nanofiber surfaces. We observe that the differences between the values of $\chi_p$ for the $x$ and $y$ polarizations [see the solid red and dashed blue lines in figure \ref{fig3}(c)] are more significant than the differences between the corresponding values of $\kappa_p$ [see the solid red and dashed blue lines in figure \ref{fig3}(a)], and the latter are in turn more significant than the differences between the corresponding values of $c_p$ [see the solid red and dashed blue lines in figure \ref{fig3}(b)]. We observe that the numerically obtained values of $\kappa_p$, $c_p$, and $\chi_p$ are positive for all values of the fiber radius $a$.

%%%%%%%%%%%%%%%%%%%%%%% Figure 4
\begin{figure}[tbh]
	\begin{center}
		\includegraphics{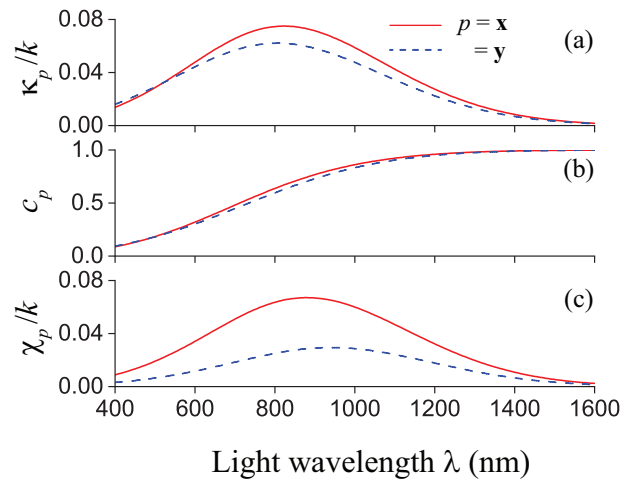}
	\end{center}
	\caption{Directional coupling coefficient $\kappa_p$ (a), butt coupling coefficient $c_p$ (b), and self-coupling coefficient $\chi_p$ as functions of the light wavelength $\lambda$. 
		The coefficients $\kappa_p$ and $\chi_p$ are normalized to the wave number $k=2\pi/\lambda$ of light. The fiber radius is $a=200$ nm and the separation distance between the fibers is $d=0$. The mode polarization is $p=\mathbf{x}$ (solid red lines) or $\mathbf{y}$ (dashed blue lines). 
		The refractive index of the nanofibers is calculated from the four-term Sellmeier formula for fused silica \cite{Malitson,Ghosh}. 		
	}
	\label{fig4}
\end{figure}

In figure \ref{fig4}, we plot the coefficients  $\kappa_p$,  $c_p$, and $\chi_p$ as functions of the light wavelength $\lambda$. 
The fiber radius is $a=200$ nm and the separation distance between the fibers is $d=0$. 
The refractive index of the nanofibers depends on the wavelength of light and is calculated from the four-term Sellmeier formula for fused silica \cite{Malitson,Ghosh}.
We observe that $\kappa_p$ and $\chi_p$ have local maxima while $c_p$ monotonically increases with increasing $\lambda$. In the limit of short wavelengths $\lambda/2\pi\ll a$, the coefficients $\kappa_p$, $c_p$, and $\chi_p$ reduce to zero. In the limit of long wavelengths $\lambda/2\pi\gg a$, the coefficients $\kappa_p$ and $\chi_p$ reduce to zero and the coefficient $c_p$ increases to 1. We observe that the numerically obtained values of $\kappa_p$, $c_p$, and $\chi_p$ are positive for all values of $\lambda$. 
Note that the fiber size parameter $ka$ determines the relative size of the system. Therefore, the curves of figure \ref{fig3} can be converted into those of figure \ref{fig4} through the transformation $\lambda=\lambda_0a_0/a$ for the horizontal axis. Here, we have the values $\lambda_0=800$ nm and $a_0=200$ nm, which were used in the calculations of figures \ref{fig3} and \ref{fig4}, respectively. This conversion is not exact because of the effect of the dispersion of the refractive index of the fiber material. However, such discrepancies are very small.

%%%%%%%%%%%%%%%%%%%%%%% Figure 5
\begin{figure}[tbh]
	\begin{center}
		\includegraphics{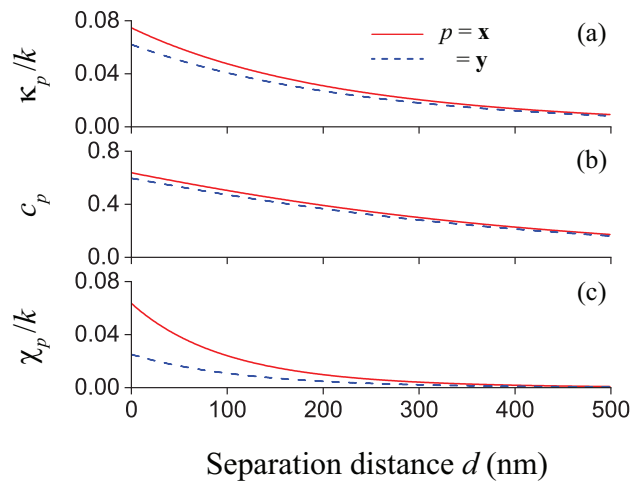}
	\end{center}
	\caption{Directional coupling coefficient $\kappa_p$ (a), butt coupling coefficient $c_p$ (b), and self-coupling coefficient $\chi_p$ as functions of the fiber separation distance $d$. The coefficients $\kappa_p$ and $\chi_p$ are normalized to the wave number $k=2\pi/\lambda$ of light. The wavelength of light is $\lambda=800$ nm and the fiber radius is $a=200$ nm. The mode polarization is $p=\mathbf{x}$ (solid red lines) or $\mathbf{y}$ (dashed blue lines). 
		Other parameters are as in figure \ref{fig2}.
	}
	\label{fig5}
\end{figure}

We plot in figure \ref{fig5} the coefficients $\kappa_p$, $c_p$, and $\chi_p$ as functions of the fiber separation distance $d$. The wavelength of light is $\lambda=800$ nm and the fiber radius is $a=200$ nm. We observe from the figure that
$\kappa_p$, $c_p$, and $\chi_p$ are significant in the region of small $d$, and reduce monotonically with increasing $d$. This reduction is a consequence of the evanescent wave nature of the guided modes outside the fibers. Comparison between figures \ref{fig5}(a), \ref{fig5}(b), and \ref{fig5}(c) shows that, when $d$ increases, $\chi_p$ and $c_p$ reduce faster and more slowly, respectively, than $\kappa_p$. It is clear from figures \ref{fig5}(a) and \ref{fig5}(c) that $\chi_p$ is comparable to or much smaller than $\kappa_p$ in the region of small or large separation distances $d$, respectively. The differences between the values of $\chi_p$ for the $x$ and $y$ polarizations [see the solid red and dashed blue lines in figure \ref{fig5}(c)] are more significant than the differences between the corresponding values of $\kappa_p$ [see the solid red and dashed blue lines in figure \ref{fig5}(a)], and the latter are in turn more significant than the differences between the corresponding values of $c_p$ [see the solid red and dashed blue lines in figure \ref{fig5}(b)]. We observe that the numerically obtained values of $\kappa_p$, $c_p$, and $\chi_p$ are positive for all values of the fiber separation distance $d$. 

According to \cite{Snyder1983,Marcus1989,Okamoto2006}, in most of the conventional analyses of directional couplers, the butt coupling coefficients $c_p$ and the self-coupling coefficient $\chi_p$ were neglected. Our numerical results presented in figures \ref{fig3}--\ref{fig5} show that, in the case of coupled nanofibers, the coefficients $c_p$ and $\chi_p$ are significant and, hence, should be taken into account. We note that, in \cite{Glorieux2019}, where the coupling of two parallel nanofibers was studied, the directional coupling coefficient was calculated but the butt coupling coefficient and the self-coupling coefficient were omitted from consideration. 

In general, an arbitrary fundamental guided mode is a superposition of the $x$- and $y$-polarized modes. Therefore, we can decompose the electric and magnetic parts of the field in an arbitrary mode $p$ of nanofiber $j$ as
$\boldsymbol{\mathcal{E}}_j^{(p)}=\alpha_{pj}^{(\mathbf{x})} \boldsymbol{\mathcal{E}}_j^{(\mathbf{x})}+\alpha_{pj}^{(\mathbf{y})} \boldsymbol{\mathcal{E}}_j^{(\mathbf{y})}$ and $\boldsymbol{\mathcal{H}}_j^{(p)}=\alpha_{pj}^{(\mathbf{x})} \boldsymbol{\mathcal{H}}_j^{(\mathbf{x})}+\alpha_{pj}^{(\mathbf{y})} \boldsymbol{\mathcal{H}}_j^{(\mathbf{y})}$,
where $\alpha_{pj}^{(\mathbf{x})}$ and $\alpha_{pj}^{(\mathbf{y})}$ are coefficients determining the overall power and polarization of the field $(\boldsymbol{\mathcal{E}}_j^{(p)},\boldsymbol{\mathcal{H}}_j^{(p)})$ through
the sum $|\alpha_{pj}^{(\mathbf{x})}|^2+|\alpha_{pj}^{(\mathbf{y})}|^2$ and the ratio $\alpha_{pj}^{(\mathbf{x})}/\alpha_{pj}^{(\mathbf{y})}$. 
For convenience and without loss of generality, we normalize the power so that $|\alpha_{pj}^{(\mathbf{x})}|^2+|\alpha_{pj}^{(\mathbf{y})}|^2=1$.
With the help of equations (\ref{a4}) and (\ref{a11}),
we can show that 
\begin{eqnarray}\label{a15a}
\kappa_{jj'}^{(pp')}&=&\alpha_{pj}^{(\mathbf{x})*}\alpha_{p'j'}^{(\mathbf{x})}\kappa_{\mathbf{x}}
+\alpha_{pj}^{(\mathbf{y})*}\alpha_{p'j'}^{(\mathbf{y})}\kappa_{\mathbf{y}},\nonumber\\
c_{jj'}^{(pp')}&=&\alpha_{pj}^{(\mathbf{x})*}\alpha_{p'j'}^{(\mathbf{x})}c_{\mathbf{x}}
+\alpha_{pj}^{(\mathbf{y})*}\alpha_{p'j'}^{(\mathbf{y})}c_{\mathbf{y}},\nonumber\\
\chi_{j}^{(pp')}&=&\alpha_{pj}^{(\mathbf{x})*}\alpha_{p'j}^{(\mathbf{x})}\chi_{\mathbf{x}}
+\alpha_{pj}^{(\mathbf{y})*}\alpha_{p'j}^{(\mathbf{y})}\chi_{\mathbf{y}}.
\end{eqnarray}
Equations (\ref{a15a}) allow us to calculate the coefficients $\kappa_{jj'}^{(pp')}$, $c_{jj'}^{(pp')}$, and $\chi_{j}^{(pp')}$ for the coupling between the modes with arbitrary polarizations $p$ and $p'$. In general, the mode expansion coefficients $\alpha_{pj,p'j'}^{(\mathbf{x,y})}$ are complex-valued, and hence so are the mode coupling coefficients
$\kappa_{jj'}^{(pp')}$, $c_{jj'}^{(pp')}$, and $\chi_{j}^{(pp')}$ although they all satisfy the Hermitian conjugate relationship, namely, $\kappa_{jj'}^{(pp')}=\kappa_{j'j}^{(p'p)*}$, $c_{jj'}^{(pp')}=c_{j'j}^{(p'p)*}$, 
and $\chi_{j}^{(pp')}=\chi_{j}^{(p'p)*}$ \cite{footnote}.

\section{Power transfer and phase shift}
\label{sec:transfer}

Equations (\ref{a8}) indicate that the mode with the principal quasilinear polarization $p=\mathbf{x}$ or $\mathbf{y}$ of nanofiber $j=1,2$ is coupled
to the mode with the same polarization $p$ of the other nanofiber $j'=2,1$ but not to the modes with the orthogonal polarization $p'=\mathbf{y}$ or $\mathbf{x}$ of both nanofibers. Due to this diagonal coupling property for the modes with the principal polarizations, the set of the coupled mode equations (\ref{a3}) splits into two independent sets, one  for $p=\mathbf{x}$ and the other one for $p=\mathbf{y}$. These independent sets for the modes with the same principal quasilinear polarization $p=\mathbf{x}$ or $\mathbf{y}$ can be written in the same form
\begin{eqnarray}\label{a15}
\frac{\rmd A_1^{(p)}}{\rmd z}+c_p\frac{\rmd A_2^{(p)}}{\rmd z}-\rmi\chi_p A_1^{(p)}-\rmi\kappa_p A_2^{(p)}&=&0,\nonumber\\
\frac{\rmd A_2^{(p)}}{\rmd z}+c_p\frac{\rmd A_1^{(p)}}{\rmd z}-\rmi\chi_p A_2^{(p)}-\rmi\kappa_p A_1^{(p)}&=&0,
\end{eqnarray}
where $\kappa_p$, $c_p$, and $\chi_p$ are real-valued coefficients.
From equations (\ref{a15}), we obtain
\begin{equation}\label{a16}
\frac{\rmd A_1^{(p)}}{\rmd z}=\rmi\delta_p A_1^{(p)}+\rmi\eta_p A_2^{(p)},\qquad
\frac{\rmd A_2^{(p)}}{\rmd z}=\rmi\delta_p A_2^{(p)}+\rmi\eta_p A_1^{(p)},
\end{equation}
where
\begin{equation}\label{a17}
\eta_p=\frac{\kappa_p-c_p\chi_p}{1-c_p^2},\qquad \delta_p=\frac{\chi_p-c_p\kappa_p}{1-c_p^2}.
\end{equation}

The eigenvalues of the coupled differential equations (\ref{a16}) are purely imaginary and are given as $\rmi(\delta_p\pm\eta_p)$.
The corresponding eigenvectors are given as $(A_1^{(p)},A_2^{(p)})=(1,\pm1)2^{-1/2}\rme^{\rmi(\delta_p\pm\eta_p)z}$. 
They represent the normal modes of the coupled nanofibers.
The propagation constants of the normal modes are $\beta+\delta_p\pm\eta_p$, shifted from
the propagation constant $\beta$ of the local modes by an amount equal to the imaginary part of the corresponding eigenvalue.

The general solution to equations (\ref{a16}) is found to be
\begin{eqnarray}\label{a18}
A_1^{(p)}(z)&=&\rme^{\rmi\delta_p z}[A_1^{(p)}(0)\cos(\eta_p z)+\rmi A_2^{(p)}(0)\sin(\eta_p z)],\nonumber\\
A_2^{(p)}(z)&=&\rme^{\rmi\delta_p z}[A_2^{(p)}(0)\cos(\eta_p z)+\rmi A_1^{(p)}(0)\sin(\eta_p z)].
\qquad
\end{eqnarray}
It is clear from equations (\ref{a18}) that $\eta_p$ characterizes the power transfer between the modes and $\delta_p$ determines the phase shift.
We have 
\begin{equation}\label{a19}
|A_1^{(p)}(z)|^2+|A_2^{(p)}(z)|^2=|A_1^{(p)}(0)|^2+|A_2^{(p)}(0)|^2. 
\end{equation}

The total power of the field in the $p$-polarized guided modes of the entire coupled-nanofiber system is given as \cite{Okamoto2006,Hardy1985,Huang1990}
\begin{equation}\label{a18a}
P^{(p)}=[|A_1^{(p)}|^2+|A_2^{(p)}|^2+2c_p\mathrm{Re}\,(A_1^{(p)}A_2^{(p)*})]P_0,
\end{equation}
where $P_0$ is the common power of the local basis modes with the mode functions $(\boldsymbol{\mathcal{E}}_j^{(p)},\boldsymbol{\mathcal{H}}_j^{(p)})$.
We can show that
\begin{equation}\label{a18b}
P^{(p)}(z)=P^{(p)}(0).
\end{equation}
Thus, the total power of the light field in the guided modes with a given principal polarization $p=\mathbf{x},\mathbf{y}$ of the two nanofibers is conserved in the propagation process,
and hence so is the total power $P=P^{(\mathbf{x})}+P^{(\mathbf{y})}$ of the guided light field of the nanofibers.
Total power conservation is a general result for coupled lossless waveguides \cite{Snyder1983,Marcus1989,Okamoto2006,Chuan1988}. The conservation of the total power of the local copropagating guided modes of the coupled fibers is a consequence of the fact that the coupling between local counterpropagating modes and the coupling between local guided and radiation modes are negligible. 

It is tricky to define the guided power in an individual waveguide.
According to the conventional coupled mode theory \cite{Hardy1985,Huang1990},  
the power of guided light in the $p$-polarized mode of the individual fiber $j$ is defined as 
\begin{equation}\label{a18c}
P_j^{(p)}(z)=|A_j^{(p)}(z)|^2P_0.
\end{equation}   
Meanwhile, according to the nonorthogonal coupled mode theory \cite{Hardy1985,Huang1990}, if one fiber, for example, fiber $j'$, is terminated at $z$, then  
the power remaining in the $p$-polarized mode of the other fiber, namely fiber $j\not=j'$, is defined as 
\begin{equation}\label{a18d}
\tilde{P}_j^{(p)}(z)=|A_j^{(p)}(z)+c_pA_{j'}^{(p)}(z)|^2P_0=|\tilde{A}_j^{(p)}(z)|^2P_0,
\end{equation}
where
\begin{eqnarray}\label{a18e}
\tilde{A}_j^{(p)}(z)&=&A_j^{(p)}(z)+c_pA_{j'}^{(p)}(z)\nonumber\\
&=&\rme^{\rmi\delta_p z}[\tilde{A}_j^{(p)}(0)\cos(\eta_p z)+\rmi \tilde{A}_{j'}^{(p)}(0)\sin(\eta_p z)]
\end{eqnarray}
with $\tilde{A}_j^{(p)}(0)=A_j^{(p)}(0)+c_pA_{j'}^{(p)}(0)$.
It is clear that expression (\ref{a18c}) in terms of the input amplitudes $A_1^{(p)}(0)$ and $A_2^{(p)}(0)$
is the same as expression (\ref{a18d}) in terms of the input amplitudes $\tilde{A}_1^{(p)}(0)$ and $\tilde{A}_2^{(p)}(0)$. Note that the definition (\ref{a18d}) reduces to the definition (\ref{a18c})
when butt coupling is neglected.

In general, we have $P_1^{(p)}+P_2^{(p)}\not=P^{(p)}$ and $\tilde{P}_1^{(p)}+\tilde{P}_2^{(p)}\not=P^{(p)}$. 
Thus, the sum of the powers of the two individual fibers may not be equal to the total power guided in the entire coupled-fiber system \cite{Hardy1985,Huang1990}. This is not a violation of the law of power conservation but a consequence of the definitions assumed for the powers in the individual fibers \cite{Huang1990}.

For the simplicity of analysis, we use the definition (\ref{a18c}). 
The power of the total guided light of nanofiber $j$ is given by $P_j=P_j^{(\mathbf{x})}+P_j^{(\mathbf{y})}=(|A_j^{(\mathbf{x})}|^2+|A_j^{(\mathbf{y})}|^2)P_0$. It follows from equation (\ref{a19}) that the sum $P_1^{(p)}+P_2^{(p)}$ of the powers of the light fields in the guided modes with a given principal polarization $p=\mathbf{x},\mathbf{y}$ of the two nanofibers is conserved in the propagation process, and hence so is the sum $P_1+P_2$ of the powers $P_1$ and $P_2$ of the guided light fields of nanofibers 1 and 2, respectively.

It is worth noting here that the strength of the power transfer is determined by the absolute value $|\eta_p|$ of the coefficient $\eta_p$ regardless of its sign.
The effective coupling length over which complete power transfer may occur is given by $L_p=\pi/2|\eta_p|$.

The numerator of the expression for the power transfer coefficient $\eta_p$ in equations (\ref{a17}) has two terms, namely
$\kappa_p$ and $-c_p\chi_p$. The term $\kappa_p$ describes the contribution of the directional coupling.
The term $-c_p\chi_p$ characterizes the contribution of the butt coupling and the self coupling. It is interesting to note that, for the parameters in the range of interest, the coefficients $\kappa_p$, $c_p$, and $\chi_p$ are positive (see figures \ref{fig3}--\ref{fig5}) and hence the terms $\kappa_p$ and $-c_p\chi_p$ have opposite signs. This means that the power transfer is a result of the competition between the directional coupling, from one side, and the butt coupling and the self coupling, from the other side.

Similarly, the numerator of the expression for the fiber-coupling-induced phase shift coefficient $\delta_p$ in equations (\ref{a17}) also has two terms, $\chi_p$ and $-c_p\kappa_p$, whose signs are opposite to each other. Thus, the fiber-coupling-induced phase shift of a guided mode is a result of the competition between the self coupling, from one side, and the butt coupling and the directional coupling, from the other side.

In general, we have $\eta_p\not=\kappa_p$ and $\delta_p\not=\chi_p$.
In the limit of large separation distances $d$, we have the relations $|c_p|\ll1$ and $|\chi_p|\ll|\kappa_p|$ \cite{Snyder1983,Marcus1989,Okamoto2006}, which lead to $\eta_p\simeq\kappa_p$ and $\delta_p\simeq0$. In the cases where the fiber radius $a$ is small, the light wavelength $\lambda$ is large, or the fiber separation distance $d$ is small, the butt coupling coefficient $c_p$ and the self-coupling coefficient $\chi_p$ cannot be neglected (see figures \ref{fig3}--\ref{fig5} and \cite{Snyder1983,Marcus1989,Okamoto2006}).

%%%%%%%%%%%%%%%%%%%%%%% Figure 6
\begin{figure}[tbh]
	\begin{center}
		\includegraphics{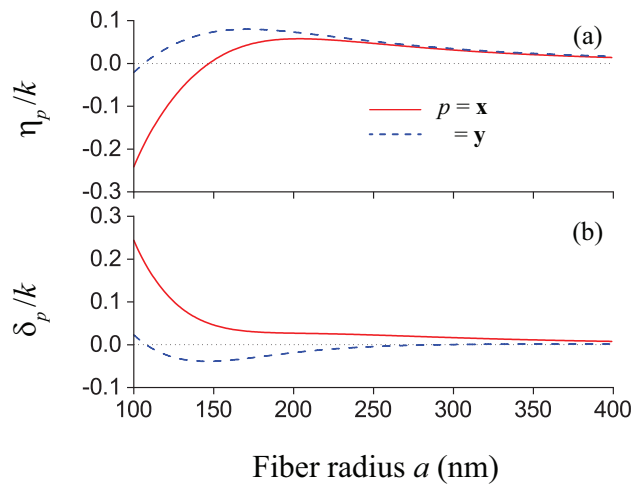}
	\end{center}
	\caption{Power transfer coefficient $\eta_p$ (a) and phase shift coefficient $\delta_p$ (b), normalized to the wave number $k$, as functions of the fiber radius $a$ in the case where the wavelength of light is $\lambda=800$ nm and the fiber separation distance is $d=0$. Parameters used are as for figure \ref{fig3}.
	}
	\label{fig6}
\end{figure} 

We plot in figure \ref{fig6} the power transfer coefficient $\eta_p$ and the phase shift coefficient $\delta_p$ as functions of the fiber radius $a$ for the parameters of figure \ref{fig3}. Figure \ref{fig6} shows that, in the limit of small fiber radii, both $\eta_p$ and $\delta_p$ tend to nonzero limiting values, unlike $\kappa_p$ and $\chi_p$, which tend to zero. Note that the corresponding asymptotic value of $\eta_p$ is opposite to that of $\delta_p$. This feature is a consequence of the fact that $c_p\to1$ in the limit of small $a$. 

We observe from figure \ref{fig6}(a) that the power transfer coefficient $\eta_p$ has a local peak at $a\simeq 204$ and 171 nm for $p=\mathbf{x}$ (solid red curve) and $\mathbf{y}$ (dashed blue curve), respectively. This figure also shows that $\eta_p$ is zero at $a\simeq 147$ and 106 nm for $p=\mathbf{x}$ and $\mathbf{y}$, respectively. It is interesting to note that $\eta_p$ is negative, that is, $\kappa_p<c_q\chi_q$, in the regions $a\leq 147$ nm and $a\leq 106$ nm for $p=\mathbf{x}$ and $\mathbf{y}$, respectively. Careful inspection of the data presented in figures \ref{fig3} and \ref{fig6}(a) shows that, in the aforementioned regions, where $\eta_p$ is negative, we have $c_p\simeq 1$ and $\kappa_p<\chi_p$. Thus, the negative sign of $\eta_p$ for small fiber radii $a$
 is a signature of the facts that the butt coupling coefficient $c_p$ is close to unity and the self-coupling coefficient $\chi_p$ is larger than the directional coupling coefficient $\kappa_p$. As already mentioned, the strength of the power transfer is determined by the absolute value $|\eta_p|$ of the coefficient $\eta_p$. We observe that, when $a$ is large enough, the effects of the butt coupling and the mode energy changes are less significant than that of the directional coupling and, hence, $\eta_p$ is positive. Comparison between the solid red and dashed blue curves of figure \ref{fig6}(a) shows that we may have all the three possibilities $|\eta_{\mathbf{x}}|<|\eta_{\mathbf{y}}|$, $|\eta_{\mathbf{x}}|=|\eta_{\mathbf{y}}|$, and $|\eta_{\mathbf{x}}|>|\eta_{\mathbf{y}}|$ depending on the fiber radius $a$. We observe from figure \ref{fig6} that the differences between $\eta_{\mathbf{x}}$ and $\eta_{\mathbf{y}}$ and between $\delta_{\mathbf{x}}$ and $\delta_{\mathbf{y}}$ are significant in the region of small $a$ but not significant in the region of large $a$.

%%%%%%%%%%%%%%%%%%%%%%% Figure 7
\begin{figure}[tbh]
	\begin{center}
		\includegraphics{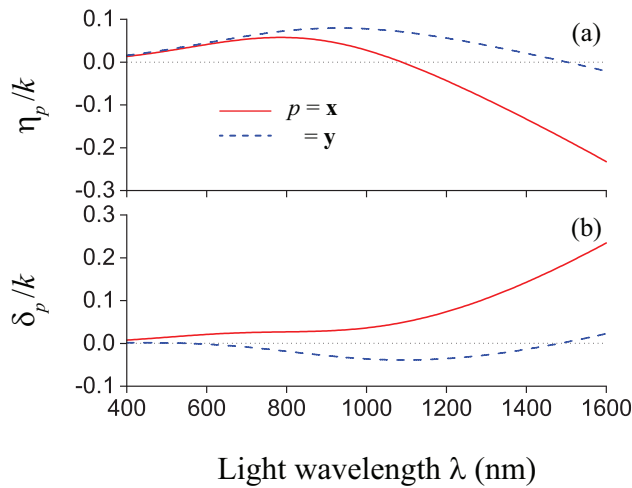}
	\end{center}
	\caption{Power transfer coefficient $\eta_p$ (a) and phase shift coefficient $\delta_p$ (b), normalized to the wave number $k$, as functions of the wavelength $\lambda$ of light in the case where the fiber radius is $a=200$ nm and the fiber separation distance is $d=0$. Parameters used are as for figure \ref{fig4}.
	}
	\label{fig7}
\end{figure}

We plot in figure \ref{fig7} the power transfer coefficient $\eta_p$ and the phase shift coefficient $\delta_p$ as functions of the wavelength $\lambda$ of light for the parameters of figure \ref{fig4}. Figure \ref{fig7} shows that, in the limit of long wavelengths, both $\eta_p$ and $\delta_p$ tend to nonzero limiting values, unlike $\kappa_p$ and $\chi_p$, which tend to zero. In this limit, since $c_p\to1$, the asymptotic value of $\eta_p$ is opposite to that of $\delta_p$.

We observe from figure \ref{fig7}(a) that the power transfer coefficient $\eta_p$ has a local peak at $\lambda\simeq 785$ and 930 nm for $p=\mathbf{x}$ (solid red curve) and $\mathbf{y}$ (dashed blue curve), respectively. The figure shows that $\eta_p$ is zero at $\lambda\simeq 1088$ and 1501 nm for $p=\mathbf{x}$ and $\mathbf{y}$, respectively, and that $\eta_p$ is negative in the regions $\lambda\geq 1088$ nm and $\lambda\geq 1501$ nm for $p=\mathbf{x}$ and $\mathbf{y}$, respectively. The facts that $\eta_p$ can become zero or negative are consequences of the competition between the contribution of the directional coupling, from one side, and the contribution of the butt coupling and the mode energy changes, from the other side. When $\lambda$ is small enough, the effects of the butt coupling and the mode energy changes are less significant than those of the directional coupling and, hence, $\eta_p$ is positive. Comparison between the solid red and dashed blue curves of figure \ref{fig7}(a) shows that we may have all the three possibilities $|\eta_{\mathbf{x}}|<|\eta_{\mathbf{y}}|$, $|\eta_{\mathbf{x}}|=|\eta_{\mathbf{y}}|$, and $|\eta_{\mathbf{x}}|>|\eta_{\mathbf{y}}|$ depending on the wavelength $\lambda$. We observe from figure \ref{fig7} that 
the differences between $\eta_{\mathbf{x}}$ and $\eta_{\mathbf{y}}$ and between $\delta_{\mathbf{x}}$ and $\delta_{\mathbf{y}}$ are significant in the region of large $\lambda$ but not significant in the region of small $\lambda$.
Since the relative size of the system is determined by the fiber size parameter $ka$, the curves of figure \ref{fig6} can be converted into those of figure \ref{fig7} through the transformation $\lambda=\lambda_0a_0/a$ for the horizontal axis, like the case of figures \ref{fig3} and \ref{fig4}. 

%%%%%%%%%%%%%%%%%%%%%%% Figure 8
\begin{figure}[tbh]
	\begin{center}
		\includegraphics{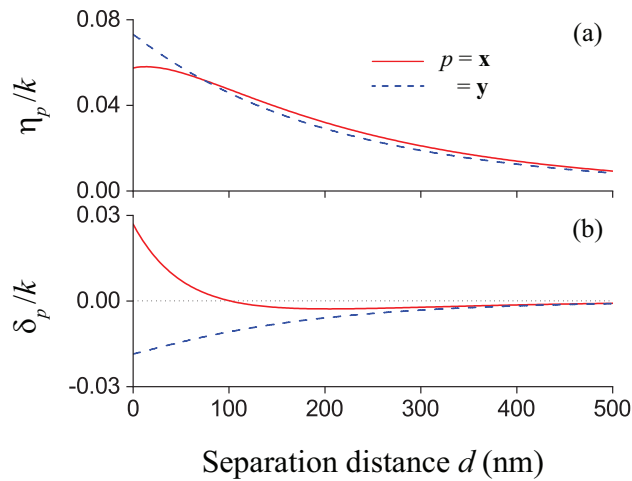}
	\end{center}
	\caption{Power transfer coefficient $\eta_p$ (a) and phase shift coefficient $\delta_p$ (b), normalized to the wave number $k$, as functions of the fiber separation distance $d$ in the case where the wavelength of light is $\lambda=800$ nm and the fiber radius is $a=200$ nm. Parameters used are as for figure \ref{fig5}.
	}
	\label{fig8}
\end{figure}

We plot in figure \ref{fig8} the power transfer coefficient $\eta_p$ and the phase shift coefficient $\delta_p$ as functions of the fiber separation distance $d$ for the parameters of figure \ref{fig5}. We observe from the solid red curve of figure \ref{fig8}(a) that the coefficient $\eta_p$ for the mode with the polarization $p=\mathbf{x}$ achieves its maximum value at a nonzero separation distance $d=14$ nm. This behavior is different from that of $\kappa_p$, which reduces monotonically with increasing $d$ [see figure \ref{fig5}(a)]. The occurrence of a local peak in the dependence of the power transfer coefficient $\eta_p$ on $d$ is a result of the competition between the directional coupling coefficient $\kappa_p$, the butt coupling coefficient $c_p$, and the self-coupling coefficient $\chi_p$ [see the expression for $\eta_p$ in equations (\ref{a17})]. We observe from figure \ref{fig8}(a) that $\eta_p$ is positive.
This feature is a signature of the fact that, when $a$ is large enough, the effect of the directional coupling on the power transfer is more significant than that of the butt coupling and the mode energy changes. Comparison between the solid red and dashed blue curves of figure \ref{fig8}(a) shows that we may have all the three possibilities $|\eta_{\mathbf{x}}|<|\eta_{\mathbf{y}}|$, $|\eta_{\mathbf{x}}|=|\eta_{\mathbf{y}}|$, and $|\eta_{\mathbf{x}}|>|\eta_{\mathbf{y}}|$ depending on the fiber separation distance $d$. In the case of figure \ref{fig8}(a), we have $|\eta_{\mathbf{x}}|<|\eta_{\mathbf{y}}|$ and $|\eta_{\mathbf{x}}|>|\eta_{\mathbf{y}}|$ in the regions $d<73$ nm and $d>73$ nm, respectively. We observe from figure \ref{fig8} that the differences between $\eta_{\mathbf{x}}$ and $\eta_{\mathbf{y}}$ and between $\delta_{\mathbf{x}}$ and $\delta_{\mathbf{y}}$ are significant in the region of small $d$ but not significant in the region of large $d$.

%%%%%%%%%%%%%%%%%%%%%%% Figure 9
\begin{figure}[tbh]
	\begin{center}
		\includegraphics{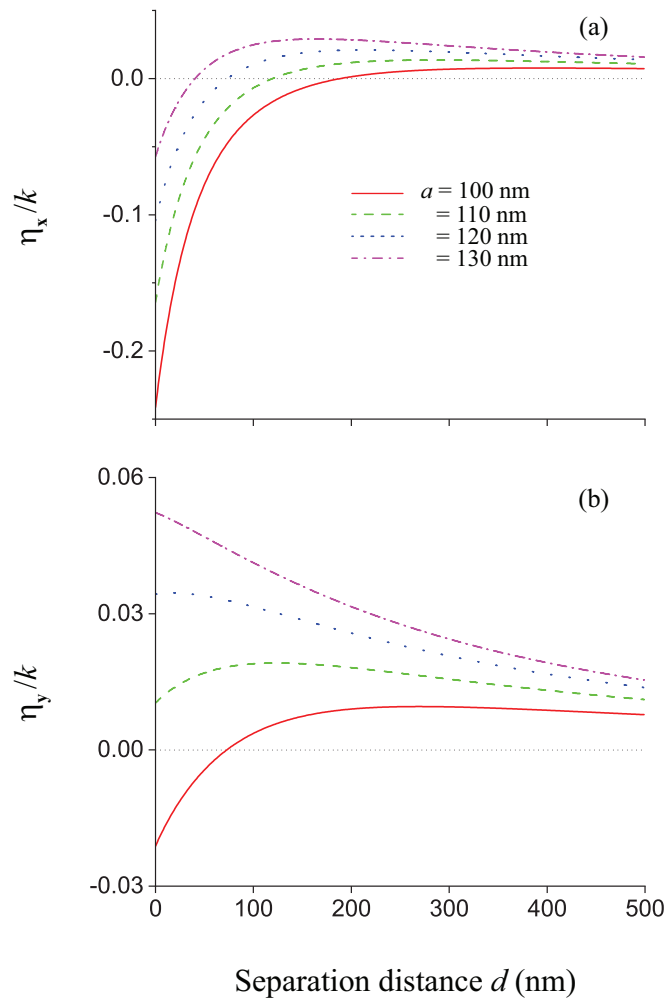}
	\end{center}
	\caption{Power transfer coefficients $\eta_{\mathbf{x}}$ (a) and $\eta_{\mathbf{y}}$ (b), normalized to the wave number $k$, as functions of the fiber separation distance $d$. The fiber radius is $a=100$ nm (solid red lines), 110 nm (dashed green lines), 120 nm (dotted blue lines), and 130 nm (dash-dotted magenta lines). 
		The wavelength of light is $\lambda=800$ nm.
		Other parameters are as in figure \ref{fig2}.
	}
	\label{fig9}
\end{figure} 

In order to see more clearly the effects of the butt coupling and the mode energy changes on the power transfer for small fiber radii and short fiber separation distances, we plot in figure \ref{fig9} the power transfer coefficients $\eta_{\mathbf{x}}$ and $\eta_{\mathbf{y}}$ as functions of the fiber separation distance $d$ for the fiber radii $a=100$, 110, 120, and 130 nm. The figure shows that, in the cases where $p=\mathbf{x}$ and $\mathbf{y}$, the coefficient $\eta_p$ for power transfer can become negative, be equal to zero, or have a local peak at a nonzero $d$. These features are consequences of the fact that the butt coupling coefficient $c_p$ and the self-coupling coefficient $\chi_p$ are significant in the regions of small fiber radii and short fiber separation distances.
We observe from figure \ref{fig9} that, when $d$ is large enough, the effects of the butt coupling and the mode energy changes on the power transfer are less significant than those of the directional coupling and, hence, $\eta_p$ is positive.

%%%%%%%%%%%%%%%%%%%%%%% Figure 10
\begin{figure}[tbh]
	\begin{center}
		\includegraphics{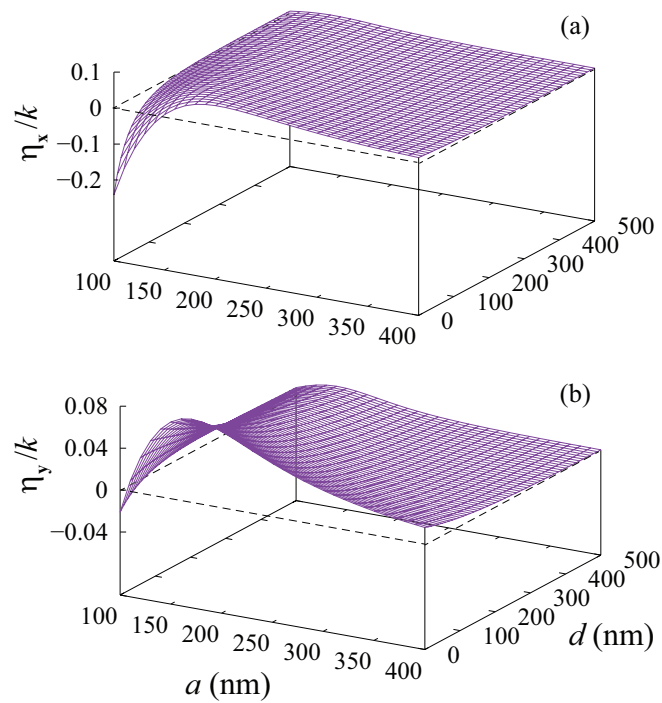}
	\end{center}
	\caption{Power transfer coefficients $\eta_{\mathbf{x}}$ (a) and $\eta_{\mathbf{y}}$ (b), normalized to the wave number $k$, as functions of the fiber radius $a$ and the fiber separation distance $d$. The wavelength of light is $\lambda=800$ nm. Other parameters are as in figure \ref{fig2}.
	}
	\label{fig10}
\end{figure}

In order to get a broader view, we plot in figure \ref{fig10} the dependencies of the power transfer coefficients $\eta_{\mathbf{x}}$ and $\eta_{\mathbf{y}}$ on the fiber radius $a$ and the fiber separation distance $d$. The figure shows clearly that, due to the effects of the butt coupling and the mode energy changes, $\eta_{\mathbf{x}}$ and $\eta_{\mathbf{y}}$ are negative when $a$ and $d$ are small enough. The coefficients $\eta_{\mathbf{x}}$ and $\eta_{\mathbf{y}}$ can become zero for some specific values of $a$ and $d$.
When $a$ or $d$ is large enough, the effects of the butt coupling and the mode energy changes on the power transfer are less significant than those of the directional coupling and, hence, $\eta_{\mathbf{x}}$ and $\eta_{\mathbf{y}}$ are positive.

%%%%%%%%%%%%%%%%%%%%%%% Figure 11
\begin{figure}[tbh]
	\begin{center}
		\includegraphics{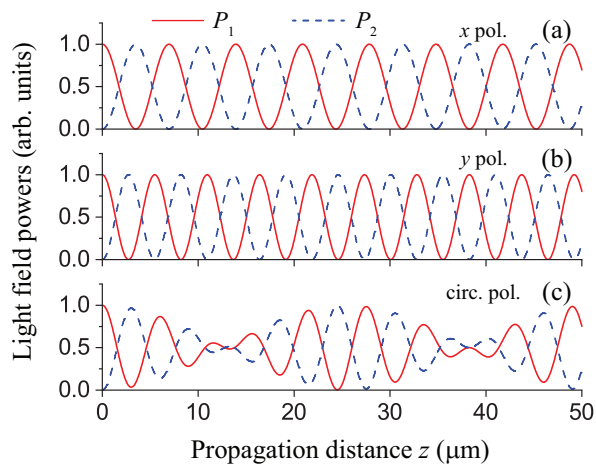}
	\end{center}
	\caption{Powers $P_1$ (solid red lines) and $P_2$ (dashed blue lines) of the guided light fields of nanofibers 1 and 2, respectively, as functions of the propagation distance $z$. The input light field for nanofiber 1 is $x$-polarized (a),
		$y$-polarized (b), or circularly polarized (c). The input light field for nanofiber 2 is zero. The fiber radius is $a=200$ nm, the wavelength of light is $\lambda=800$ nm, and the fiber separation distance is $d=0$. Other parameters are as in figure \ref{fig2}.
	}
	\label{fig11}
\end{figure}

The power transfer between the guided modes of the nanofibers is described by 
equations (\ref{a18}). We use these equations to calculate the power $P_1=(|A_1^{(\mathbf{x})}|^2+|A_1^{(\mathbf{y})}|^2)P_0$ and $P_2=(|A_2^{(\mathbf{x})}|^2+|A_2^{(\mathbf{y})}|^2)P_0$ of the guided light fields of nanofibers 1 and 2, respectively, as functions of the propagation distance $z$. We plot in figure \ref{fig11} the results of calculations for the cases where
the input field for nanofiber 1 is $x$-polarized, $y$-polarized, or circularly polarized, and the input field for nanofiber 2 is zero. 

We observe from the figure that, in the cases where the input light is quasilinearly polarized along the principal axis $x$ [see figure \ref{fig11}(a)] or $y$ [see figure \ref{fig11}(b)], the powers $P_1$ (solid red lines) and $P_2$ (dashed blue lines) vary periodically in space and 100\% power transfers occur at integer multiples of the coupling length $L_{\mathbf{x}}=\pi/2|\eta_{\mathbf{x}}|\simeq 3.48$ $\mu$m or $L_{\mathbf{y}}=\pi/2|\eta_{\mathbf{y}}|\simeq 2.73$ $\mu$m. It is clear that, in the case of figure \ref{fig11}, where $a=200$ nm and $d=0$, we obtain $L_{\mathbf{x}}>L_{\mathbf{y}}$. This feature is a consequence of the fact that, for the parameters of figure \ref{fig11}, we have $|\eta_{\mathbf{x}}|<|\eta_{\mathbf{y}}|$ [see figures \ref{fig6}(a) and \ref{fig8}(a)]. It follows from the numerical results for $\eta_p$ presented in figures \ref{fig6}(a) and \ref{fig8}(a) that we may observe all the three possibilities $L_{\mathbf{x}}<L_{\mathbf{y}}$, $L_{\mathbf{x}}=L_{\mathbf{y}}$, and $L_{\mathbf{x}}>L_{\mathbf{y}}$, depending on the fiber radius $a$, the light wavelength $\lambda$, and the fiber separation distance $d$.

We see from figure \ref{fig11}(c) that, in the case where the input light is quasicircularly polarized, the spatial dependencies of the powers $P_1$ and $P_2$ show the beating of two different harmonic waves with different wavelengths corresponding to the distinct coupling lengths $L_{\mathbf{x}}$ and $L_{\mathbf{y}}$. Near 100\% power transfers occur at specific propagation lengths.

\section{Summary}
\label{sec:summary}

In conclusion, we have studied analytically and numerically the coupling between the fundamental guided modes of two identical parallel nanofibers. We have calculated the coefficients of directional coupling, butt coupling, and mode energy changes as functions of the fiber radius, the light wavelength, and the fiber separation distance. We have shown that, due to the symmetry of the coupled fiber system, the coupling between two modes with different orthogonal principal quasilinear polarizations, aligned along the radial axis $x$ and the tangential axis $y$, vanishes. Due to this property, the $x$- and $y$-polarized guided modes of a nanofiber are coupled only to the modes with the same corresponding principal polarization of the other nanofiber. 
We have found that the effects of butt coupling and mode energy changes on the power transfer are significant when the fiber radius is small (the region $a\lesssim 200$ nm of figures \ref{fig3} and \ref{fig6}), the light wavelength is large (the region $\lambda\gtrsim 800$ nm of figures \ref{fig4} and \ref{fig7}), or the fiber separation distance is small (the region $d\lesssim 200$ nm of figures \ref{fig5} and \ref{fig8}).
We have shown that the power transfer coefficient may achieve a local maximum or become zero when the fiber radius, the light wavelength, or the fiber separation distance varies. 

We recognize that the effects of difference in radius between nanofibers on the coupling between them were not considered in our paper. For dissimilar fibers, the directional coupling coefficients $\kappa_{12}^{(p_1p_2)}$ and $\kappa_{21}^{(p_2p_1)}$ are not related by the complex conjugate relationship \cite{Okamoto2006} and, moreover, one of the fibers can couple efficiently into the other fiber but not conversely \cite{Hardy1985}. In addition, a difference in radius between nanofibers leads to a difference in propagation constant and hence to a phase mismatch causing a reduction of the power transfer between the fiber modes. 
The effects of difference in radius between nanofibers on the coupling between them deserve a separate systematic research.

Our results are important for controlling and manipulating guided fields of coupled parallel nanofibers and also have relevance in relation to nanofiber coupled ring resonators made from tapered fibers. They can be envisioned to have significant influence on ongoing and future experiments in nanofiber quantum optics and nanophotonics. 
Due to the small sizes and the high coupling efficiencies, coupled parallel nanofibers could be use as miniaturized Sagnac interferometer \cite{Glorieux2019,Ma2017} and Fabry-Perot resonators in photonic circuits \cite{Glorieux2019}. They can also be extended to be used with emitters and scatterers. 

\ack
This work was supported by the Okinawa Institute of Science and Technology Graduate University.

%%%%%%%%%%%%%%%%%%%%%%%%%%%%%%%%%%%%%%%%%%
%%%%%%%%%%%%%%%%%%%%%%%%%%%%%%%%%%%%%%%%%%

\appendix

\section{Mode functions}
\label{appendix}

According to \cite{Snyder1983,Marcus1989,Okamoto2006,Tong2004,fibermode,highorder}, the mode functions of $x$- and $y$-polarized
forward-propagating fundamental modes are given as
\begin{eqnarray}\label{b1}
\boldsymbol{\mathcal{E}}^{(\mathbf{x})}_j&=&\sqrt2[\hat{\mathbf{r}}_je_r(r_j)\cos \varphi_j+\rmi\hat{\boldsymbol{\varphi}}_je_\varphi(r_j)\sin\varphi_j
%\nonumber\\ &&\mbox{} 
+\hat{\mathbf{z}}_je_z(r_j)\cos\varphi_j],
\nonumber\\
\boldsymbol{\mathcal{H}}^{(\mathbf{x})}_j&=&\sqrt2[\rmi\hat{\mathbf{r}}_jh_r(r_j)\sin \varphi_j+\hat{\boldsymbol{\varphi}}_jh_\varphi(r_j)\cos\varphi_j
%\nonumber\\ &&\mbox{} 
+\rmi\hat{\mathbf{z}}_jh_z(r_j)\sin\varphi_j],
\end{eqnarray}
and
\begin{eqnarray}\label{b2}
\boldsymbol{\mathcal{E}}^{(\mathbf{y})}_j&=&\sqrt2[\hat{\mathbf{r}}_je_r(r_j)\sin \varphi_j-\rmi\hat{\boldsymbol{\varphi}}_je_\varphi(r_j)\cos\varphi_j
%\nonumber\\&&\mbox{} 
+\hat{\mathbf{z}}_je_z(r_j)\sin\varphi_j],
\nonumber\\
\boldsymbol{\mathcal{H}}^{(\mathbf{y})}_j&=&\sqrt2[-\rmi\hat{\mathbf{r}}_jh_r(r_j)\cos \varphi_j+\hat{\boldsymbol{\varphi}}_jh_\varphi(r_j)\sin\varphi_j
%\nonumber\\&&\mbox{} 
-\rmi\hat{\mathbf{z}}_jh_z(r_j)\cos\varphi_j],
\end{eqnarray}
where $e_r$, $e_\varphi$, and $e_z$ are the cylindrical components of quasicircularly forward-propagating fundamental modes \cite{Snyder1983,Marcus1989,Okamoto2006,Tong2004,fibermode,highorder}.
It follows from equations (\ref{b1}) and (\ref{b2}) that the Cartesian components of the mode functions are
\begin{eqnarray}\label{b3}
\mathcal{E}^{(\mathbf{x})}_{jx}&=&\sqrt2[e_r(r_j)\cos^2\varphi_j-\rmi e_\varphi(r_j)\sin^2\varphi_j],\nonumber\\
\mathcal{E}^{(\mathbf{x})}_{jy}&=&\sqrt2[e_r(r_j)+\rmi e_\varphi(r_j)]\sin\varphi_j\cos\varphi_j,\nonumber\\
\mathcal{E}^{(\mathbf{x})}_{jz}&=&\sqrt2 e_z(r_j)\cos\varphi_j,
\nonumber\\
\mathcal{H}^{(\mathbf{x})}_{jx}&=&\sqrt2[\rmi h_r(r_j)-h_\varphi(r_j)]\sin\varphi_j\cos\varphi_j,\nonumber\\
\mathcal{H}^{(\mathbf{x})}_{jy}&=&\sqrt2[\rmi h_r(r_j)\sin^2\varphi_j+h_\varphi(r_j)\cos^2\varphi_j],\nonumber\\
\mathcal{H}^{(\mathbf{x})}_{jz}&=&\sqrt2\; \rmi h_z(r_j)\sin\varphi_j,
\end{eqnarray}
and
\begin{eqnarray}\label{b4}
\mathcal{E}^{(\mathbf{y})}_{jx}&=&\sqrt2[e_r(r_j)+\rmi e_\varphi(r_j)]\sin\varphi_j\cos\varphi_j,\nonumber\\
\mathcal{E}^{(\mathbf{y})}_{jy}&=&\sqrt2[e_r(r_j)\sin^2\varphi_j-\rmi e_\varphi(r_j)\cos^2\varphi_j],\nonumber\\
\mathcal{E}^{(\mathbf{y})}_{jz}&=&\sqrt2 e_z(r_j)\sin\varphi_j,
\nonumber\\
\mathcal{H}^{(\mathbf{y})}_{jx}&=&\sqrt2[-\rmi h_r(r_j)\cos^2\varphi_j-h_\varphi(r_j)\sin^2\varphi_j],\nonumber\\
\mathcal{H}^{(\mathbf{y})}_{jy}&=&\sqrt2[-\rmi h_r(r_j)+h_\varphi(r_j)]\sin\varphi_j\cos\varphi_j,\nonumber\\
\mathcal{H}^{(\mathbf{y})}_{jz}&=&-\sqrt2\; \rmi h_z(r_j)\cos\varphi_j,
\end{eqnarray}

\end{document}